# Small simple impact craters

**Amelia Carolina Sparavigna**
Dipartimento di Fisica, Politecnico di Torino
C.so Duca degli Abruzzi 24, Torino, Italy

**Abstract**
The paper discusses some examples of image processing applied to improve optical satellite imagery of small craters (Kamil, Veevers, Haviland). The examples show that image processing can be quite useful for further in-situ researches, because the resultant imagery helps to have a better picture of the crater shape and of the distribution of debris about it. The paper is also disclosing an interesting underwater structure, with shape and size of a small crater, located on the coast-line of Sudan.

The term Crater (κρατηρ) in Greek means the bowl used to mix wine and water during symposia. This term passed to describe volcanic and impact depressions found on the surface of Earth or other rocky planets. A simple impact crater, created by the collision of a relatively small body with the surface of the planet, has a landform with roughly circular outline and a raised rim, when diameter is small, that is, the form of a bowl.
Small impact events are producing simple craters. Here we consider as small craters, those with a diameter less than 100 m. These structures are common and detectable on the rocky bodies of the Solar System. Easy to observe, these craters remain well-preserved on the lunar surface, providing a record of the rate of impacts and of the distribution of bullet sizes: to deduce a similar curve from impacts over Earth's surface is complicated [1]. Besides the fact that the atmosphere is disrupting meteoroids, and craters are removed by erosion and tectonics, or filled by sedimentation, they can remain simply unnoticed. A combination of these factors gives rise to the recognized departure from a simple power-law size distribution for terrestrial craters with a diameter less than 20 km [2]. The actual terrestrial small crater record seems, then, to be inadequate for evaluating the flux of extraterrestrial material on the Earth's surface.
The database of craters can be greatly improved by inspection of terrain maps obtained from satellite imagery, directly or by means of processing methods able to detect round circular features in the image frames. Maps can be recorded by satellites equipped with optical detectors or with remote sensing X-band SAR detectors, able to overcome the limits of optical sensors, and see the surface features under water or sand. In the case of a small crater, the Whitecourt Meteorite Impact Crater, WMIC, it was a LIDAR inspection to depict its shape removing the mask of the vegetation hiding the crater in an optical satellite imagery [3].
Here we propose some examples of a processing applied to improve the optical images of small craters (Kamil, Veevers, Haviland), based on a method already used for larger structure [4,5], and discussed in Ref.6. These examples show that the processing method can be a fundamental help for further in-situ studies of the impact area. It is able to give a better picture of the crater shape and the debris distribution about it, created by the ejecta blanket.
Veevers and Haviland are two craters from a list in Ref.7: name and position of those craters with diameter less than 100 m are shown in Table I. The satellite images are obtained from Google Maps. The table is also containing the recently discovered Kamil Crater. This is a perfectly preserved crater, observed during a survey of satellite images on Google Earth, by Vincenzo de Michele, Museo Civico di Storia Naturale, Milano, Italy. The crater was formed within the past thousand years: it is probably less than 5,000 years old [8].
As Fig.1 is showing, the Kamil crater possesses a well-defined rayed structure. The crater is shielded by some rocky hills, preserving it to be covered by the sand conveyed by the blowing of desert winds. In the right panel of Fig.1, we can see the result of processing to enhance details and adjust contrast and brightness, as made in Ref.[4,5].

Another beautiful crater is Veevers (see Fig.2), in Australia. Veevers crater is located on a flat desert plain and the site is very remote and difficult to reach. Discovered in July 1989 and named in honour of J.J. Veevers, its impact origin was confirmed in 1990 [9].

The smallest crater in Table I is the Haviland Crater. Also called the Brenham Crater, it is a meteorite crater in Kiowa County, Kansas [10]. The oval-shaped crater is roughly 15 meters. Its age is estimated to be less than 1000 years, placing it in the Holocene. It is very small in a well-preserved green meadow, quite interesting to observed with Google Maps, because it is surrounded by tilled fields. The left panel of the figure is showing as the crater appears in the map. In the middle, the same image after processing. Note the improvement of the ejecta blanket visibility after processing.

Let us conclude with the processing of an interesting underwater structure, with the size of a small crater, located on the coast-line of Sudan, see Fig.4. The presence of this object was observed during the survey in preparing Ref.5. In fact, the shape seems that of a bowl, which is bright inside for a sedimentation of sand. To the author's knowledge, this could be the first example of a small crater in shallow waters. The existence of such craters seems to be possible, because the mechanism for their formation is the same for craters in air. It is interesting to note that under certain conditions, an underwater crater can be observed within the optical range detection too.

In fact, the image processing is able to highly increase the details of impact structures, but, only a direct inspection can tell whether this is a true impact crater or not. In any case, the examples we discussed demonstrate that the database of small craters and related imagery can be greatly improved using the inspection of satellite maps combined with image processing methods.

**References**
1. P.A. Bland and N.A. Artemieva, The impact rate of small asteroids at the Earth's surface. Large Meteorite Impacts, 2003, 4047.PDF.
2. R.A.F. Grieve and E.M. Shoemaker, in Hazards due to Comets and Asteroids, T. Gehrels, Ed. (Univ. Arizona Press, Tucson), 1994, p.417.
3. R.S. Kofman, C.D.K. Herd, E.L. Walton and D.G. Froese, The late Holocene Whitecourt meteorite impact crater: a low-energy hypervelocity event. 40th Lunar and Planetary Science Conference, 2009, 1942.PDF.
4. A.C. Sparavigna, Crater-like landform in Bayuda desert (a processing of satellite images), 2010, arXiv:1008.0500, physics.geo-ph.
5. A.C. Sparavigna, Craters and ring complexes of the North-East Sudanese country 2010, arXiv:1008.3976, physics.geo-ph.
6. R. Marazzato and A.C. Sparavigna, Astronomical image processing based on fractional calculus: the AstroFracTool, 2009, arXiv:0910.4637, astro-ph.IM.
7. Earth Impact Database, www.unb.ca/passc/ImpactDatabase/index.html
8. L. Folco, M. Di Martino, A. El Barkooky, M. D'Orazio, A. Lethy, S. Urbini,. I. Nicolosi,.M. Hafez, C. Cordier, M. van Ginneken, A. Zeoli, A.M. Radwan, S. El Khrepy, M. El Gabry, M. Gomaa,.A. A. Barakat, R. Serra, M. El Sharkawi, The Kamil Crater in Egypt, Science, 22 July 2010, DOI: 10.1126/science.1190990
9. E.M. Shoemaker and C.S. Shoemaker, Impact structures of Western Australia, Meteoritics, 1985, Vol.20, pp. 754–6.
10. H.H. Nininger and J.D. Figgins, The excavation of a meteorite crater near Haviland, Kansas, American Journal of Science, Vol.28(16), 1934, p. 312-313.

Table I

| Name | Country | Coordinates | Diameter (km) |
|---|---|---|---|
| Morasko | Poland | N 52° 29', E 16° 54' | 0.1 |
| Ilumetsä | Estonia | N 57° 58', E 27° 25' | 0.08 |
| **Veevers** | Western Australia | S 22° 58', E 125° 22' | 0.08 |
| Sobolev | Russia | N 46° 18', E 137° 52' | 0.053 |
| Campo Del Cielo | Argentina | S 27° 38', W 61° 42' | 0.05 |
| Whitecourt | Alberta, Canada | N 54° 00', W 115° 36' | 0.036 |
| Sikhote Alin | Russia | N 46° 7', E 134° 40' | 0.027 |
| Dalgaranga | Western Australia | S 27° 38', E 117° 17' | 0.024 |
| **Haviland** | Kansas, U.S.A. | N 37° 35', W 99° 10' | 0.015 |
| **Kamil** | Egypt | N 22°.02011, E 26°.09059 | 0.045 |

List of craters with diameter less than 100 m, from Ref.7, with the recently discovered Kamil Crater was enclosed.

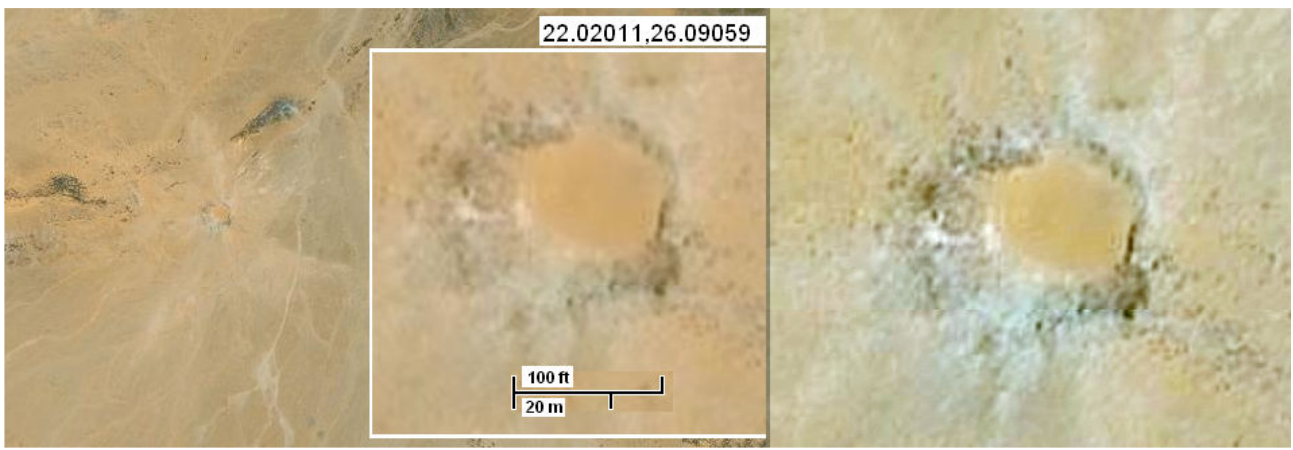

Fig.1. The Kamil Crater as observed with Google Maps. It is a perfectly preserved crater formed within the past thousand years [8]. The crater possesses a rayed structure. In the right panel, the image after processing to enhance details.

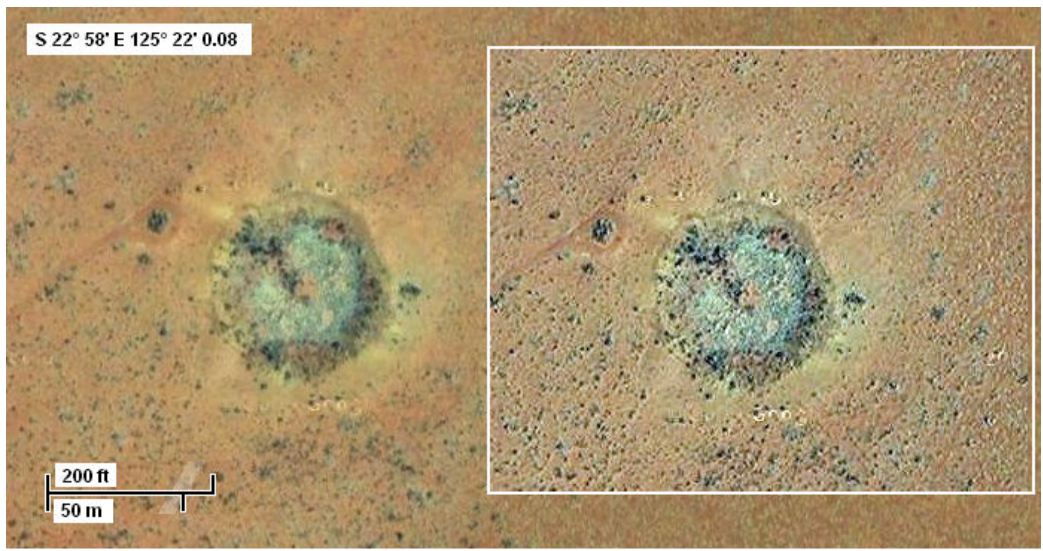

Fig.2. The Veevers Crater as observed with Google Maps. It is another well-preserved crater in Australia. In the right panel, the result after image processing.

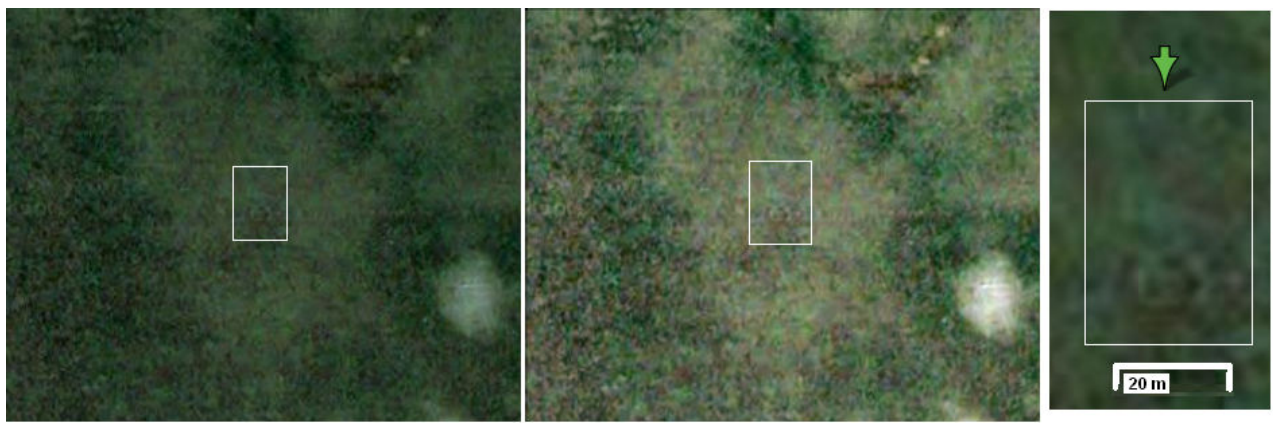

Fig.3. The smallest crater in Table I is the Haviland Crater. It is a meteorite crater in Kiowa County, Kansas. The diameter is roughly 15 meters. The left panel shows as it appears in the Google Maps. In the middle, the same image after processing. Note the improvement of ejecta blanket visibility.

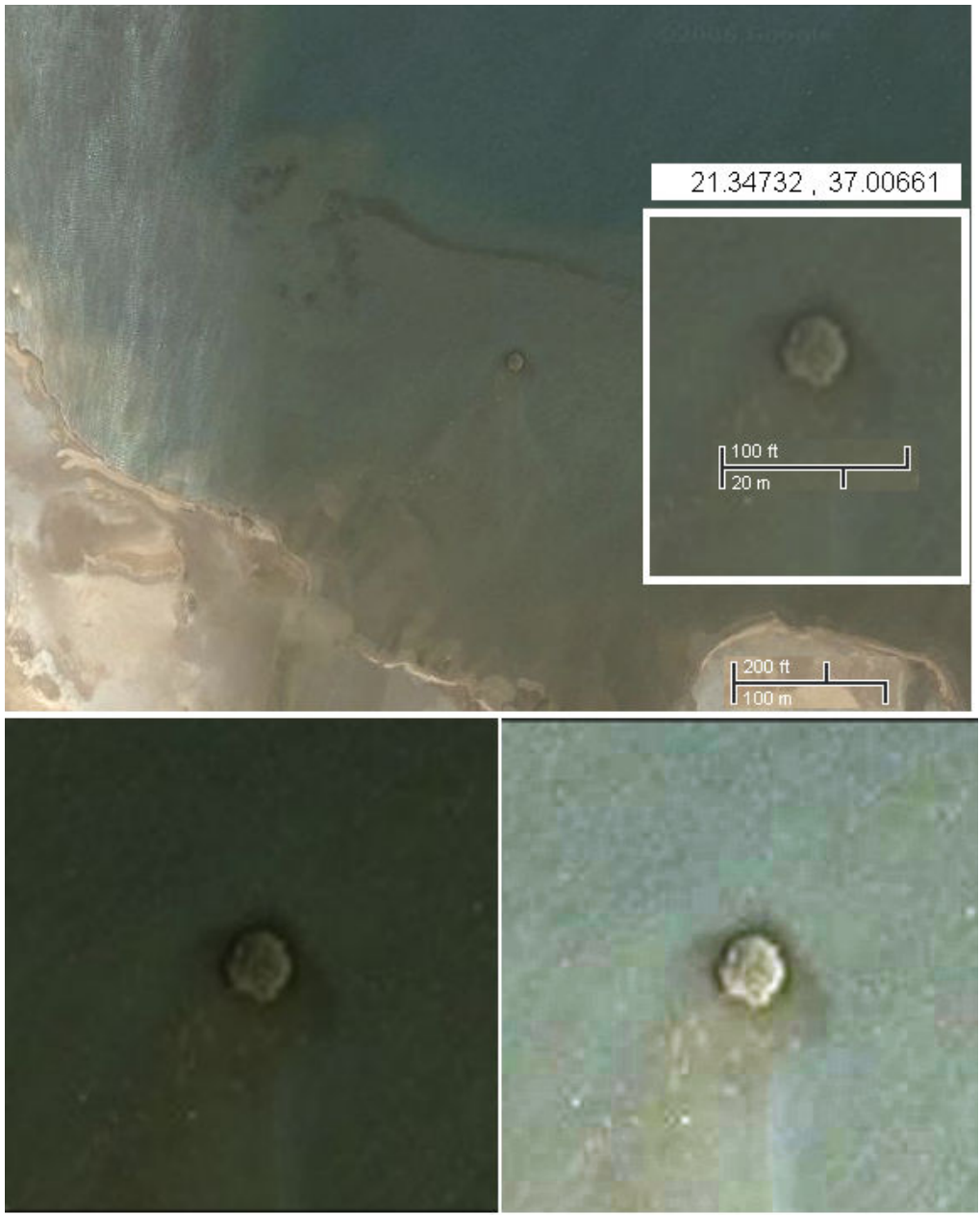

Fig.4. An interesting underwater structure of the coast-line of Sudan. Coordinates are given in the insert. In the lower part of the figure we see, on the left, the original image from Google Maps, and, on the right, the image after processing.